\documentclass{iau}
\usepackage{graphicx}
\usepackage{sidecap}
\usepackage{natbib}
\graphicspath{{./fig/}{./png/}}

\newcommand{\EQ}{\begin{equation}}
\newcommand{\EN}{\end{equation}}
\newcommand{\EQA}{\begin{eqnarray}}
\newcommand{\ENA}{\end{eqnarray}}

\newcommand{\Eq}[1]{Equation~(\ref{#1})}

\newcommand{\Fig}[1]{Figure~\ref{#1}}

\newcommand{\bra}[1]{\langle #1\rangle}

\newcommand{\mean}[1]{\overline #1}
\newcommand{\meanrho}{\overline{\rho}}

{}
{}
{}

{}
{}
{}
{}
{}
{}
{}
{}
{}
\newcommand{\meanBB}{\overline{\mbox{\boldmath $B$}}{}}{}
{}
{}
{}
{}
{}
{}
{}
{}

{}
{}
{}

{}

{}
{}

\newcommand{\gggg}{\mbox{\boldmath $g$} {}}

\newcommand{\rr}{\mbox{\boldmath $r$} {}}

\newcommand{\uu}{\mbox{\boldmath $u$} {}}

\newcommand{\bb}{\mbox{\boldmath $b$} {}}
\newcommand{\BB}{\mbox{\boldmath $B$} {}}

\newcommand{\jj}{\mbox{\boldmath $j$} {}}
\newcommand{\JJ}{\mbox{\boldmath $J$} {}}

\newcommand{\AAA}{\mbox{\boldmath $A$} {}}

\newcommand{\nab}{\mbox{\boldmath $\nabla$} {}}
\newcommand{\OO}{\mbox{\boldmath $\Omega$} {}}
\newcommand{\oo}{\mbox{\boldmath $\omega$} {}}

\newcommand{\chit}{\chi_{\rm t}}

\newcommand{\SSSS}{\mbox{\boldmath ${\sf S}$} {}}

\newcommand{\DD}{{\rm D} {}}

\def\Co{\mbox{\rm Co}}

\def\Pm{\mbox{\rm Pr}_M}

\def\Rey{\mbox{\rm Re}}

\def\Co{\mbox{\rm Co}}

\def\cs{c_{\rm s}}

\def\kf{k_{\rm f}}

\def\Brms{B_{\rm rms}}

\def\urms{u_{\rm rms}}

\def\Beq{B_{\rm eq}}

\def\onethird{{\textstyle{1\over3}}}

\newcommand{\yapj}[3]{ #1, {ApJ,} {#2}, #3}

\newcommand{\yapjl}[3]{ #1, {ApJL,} {#2}, #3}
\newcommand{\yapjs}[3]{ #1, {ApJS,} {#2}, #3}

\newcommand{\yan}[3]{ #1, {Astron.\ Nachr.,} {#2}, #3}

\newcommand{\yana}[3]{ #1, {A\&A,} {#2}, #3}

\newcommand{\ysph}[3]{ #1, {Solar Phys.,} {#2}, #3}
\newcommand{\yswsc}[3]{ #1, {JSWSC,} {#2}, #3}

\title[Solar-like differential rotation and equatorward migration] 
{Solar-like differential rotation and equatorward migration in a
  convective dynamo with a coronal envelope}

\author[Warnecke et al.]   
{J.\ Warnecke$^{1,2}$
 P.\ J.\ K{\"a}pyl{\"a}$^{1,3}$
 M.\ J.\ Mantere$^{3}$
 \and A.\ Brandenburg$^{1,2}$}
\affiliation{$^1$NORDITA, KTH Royal Institute of Technology and Stockholm University,
Roslagstullsbacken 23, SE-10691 Stockholm, Sweden, email: {\tt joern@nordita.org} \\
$^2$Department of Astronomy, Stockholm University,
SE-10691 Stockholm, Sweden\\
$^{3}$ Department of Physics, PO BOX 64, FI-00014 Helsinki University, Finland}

\pubyear{2013}
\volume{294}  
\pagerange{1--5}
\jname{Solar and Astrophysical Dynamos and Magnetic Activity}
\editors{A.G. Kosovichev, E.M. de Gouveia Dal Pino, \& Y.Yan, eds.}
\doi{}

\begin{document}

\maketitle

\begin{abstract}
We present results of convective turbulent dynamo simulations
including a coronal layer in a spherical wedge.
We find an equatorward migration of the radial and azimuthal fields
similar to the behavior of sunspots during the solar cycle.
The migration of the field coexist with a spoke-like differential
rotation and anti-solar (clockwise) meridional circulation.
Even though the migration extends over the whole convection zone, the
mechanism causing this is not yet fully understood.

\end{abstract}
\firstsection 
\section{Introduction}
The Sun shows an activity cycle with a period of eleven years, which is
manifested by sunspots occurring on the solar disk.
Every eleven years the number of sunspots has a maximum and around five
years later a
minimum.
With each cycle the polarity of the global magnetic field changes
sign and gives rise to the 22 years {\it Hale} cycle.
The occurrence of sunspots does not only vary with time, but it also
shows a strong latitudinal dependence.
During solar minimum, the sunspots are more likely to emerge at higher
latitudes and during the maximum at lower latitudes, close to the
equator.
This observed feature corresponds to an equatorward migration of the
magnetic activity belt causing the sunspots.
Producing the cyclic behavior in numerical simulations is
challenging.
Either one neglects important backreaction of the magnetic field on
the fluid motions and solves the problem in a kinematic mean-field
approach with widely varying parameterizations for the small scales
\citep[e.g.][]{DC99,KKT06,KO12} or one uses direct numerical
simulations (DNS) with
orders of magnitudes too high viscosity and diffusivity, and too low
stratification of density and temperature
\citep[e.g.][]{BMT04,GCS10,KKBMT10}.
The kinematic mean-field approach seems to be successful in reproducing
observed features of the Sun. However, whether solar activity is
generated by a distributed turbulent dynamo throughout the convection
zone or in a thin layer below the convection zone is still very much
under debate and cannot be answered just by mean-field modeling.
It has turned out to be much more challenging to obtain solar-like
magnetic activity from DNS or large-eddy simulations: models either
show only weak mean fields \citep{BMT04}, strong quasi-steady
large-scale fields \citep{BBBMT10}, cycles but no migration
\citep{GCS10}, or cycles and poleward migration
\citep{G83,KKBMT10,BMBBT11}.
Recently \cite{KMB12} were able to reproduce an equatorward migration
of the
azimuthal and radial magnetic fields for the first time using direct
numerical simulation of a convective dynamo.

Another important observed feature of the Sun is its differential
rotation.
It is broadly believed that the solar differential rotation is an 
important
ingredient in the solar dynamo.
In the kinematic models the differential rotation profile is used as
an input parameter.
Beside mean-field models \citep[e.g.][]{KR99}, a
solar-like differential rotation profile
has not yet been self-consistently obtained in direct numerical 
simulations.
Often the rotation profile is solar-like in the sense that the equator
rotates faster than the poles \citep[e.g.][]{BMT04,KMB11}, but the
angular velocity is constant on cylinders due to the Taylor-Proudman
balance. {\it Spoke-like} rotation has not been seen unless a
latitudinal temperature gradient is imposed at the base of the
convection zone \citep{MBT06}.

In the following work, we present the results of a two-layer model,
which combines a convection zone with a coronal layer.
The model was originally developed to investigate coronal ejections
generated by a dynamo underneath the surface and the helicity flux due
to the coronal envelope and the ejections, see
\cite{WB10} and \cite{WBM11,WBM12,WKMB12} for details.
In this work we focus on the magnetic and fluid properties in the
convection zone.
The dynamics of coronal ejections will be discussed in a followup paper.
\section{The model}
\label{model}
\begin{figure}[t!]
\begin{center}
\includegraphics[width=0.7\columnwidth]{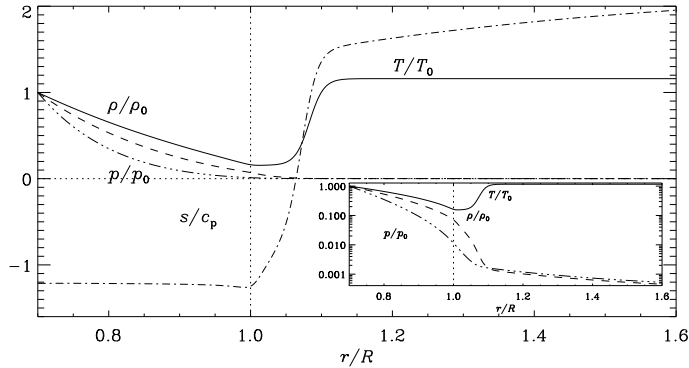}
\end{center}\caption[]{
Overview of the stratification.
The normalized density $\rho/\rho_0$, pressure $p/p_0$,
temperature $T/T_0$ are plotted together with the
specific entropy $s/c_p$ over the radius.
The inset in logarithmic scale emphasizes the steep decrease of 
the pressure and density in the coronal layer.
}
\label{strat}
\end{figure}
We use a similar setup as in \cite{WKMB12}, which we improve in two
respects.
Firstly, we apply more solar-like temperature and density profiles 
for the
coronal layer.
In the earlier work, the temperature in the corona was constant and
identical with that in the photosphere.
Now, we let the temperature rise above the surface reaching a constant
value at a radius of $r=1.1\, R$, where $R$ is the solar radius.
The coronal temperature is 1.2 times larger than that at the base 
of the
convection zone.
In \Fig{strat} the radial profiles of density, temperature, pressure,
and entropy are shown from the saturated stage of the simulation.
The entropy is negative in the convection zone with a negative slope.
Close to the surface ($r=R$) the gradient changes sign resulting in
positive values above $r=1.05\, R$.
The density decreases exponentially in the corona ($r>R$). In the
convection zone ($0.7R<r<R$) the stratification is initially
isentropic. This results in a density difference $\rho_0/\rho_{\rm s}=14$ in
the convection zone and $\rho_0/\rho_{\rm t}=2.3\cdot10^3$ in the entire
domain,
where $\rho_0=\rho(r=0.7\, R)$, $\rho_{\rm s}=\rho(r=R)$ and
$\rho_{\rm t}=\rho(r=1.6\, R)$.
The other improvement is to use a turbulent heat conductivity $\chit$
in the entropy equation (\ref{entr}) to model the unresolved heat
conductivity.
Similar subgrid-scale models have also been used by e.g.\
\cite{BMT04} and \cite{KMB11} to significantly increase energy 
transport
due to the convective flux.
We apply a constant value of $\chit$, which is similar to the
molecular viscosity $\nu$, in the interval of $0.75R\leq r\leq
0.97R$, and tends smoothly to zero above and below.

We use a spherical wedge extending from $r=0.7\, R$ to $r=1.6\, R$ in
radial direction, from $\theta=15^{\circ}$ to $165^{\circ}$ in
colatitude, and from $\phi=0^{\circ}$ to $90^{\circ}$ in azimuthal
direction.
We solve the following
equations of compressible magnetohydrodynamics:
\begin{eqnarray}
{\partial\AAA\over\partial t}&=&\uu\times\BB+\eta\nab^2\AAA,\\
{\DD\ln\rho\over \DD t} &=&-\nab\cdot\uu,\\
{\DD\uu\over\DD t}&=&  \gggg - 2\OO \times \uu + {1\over\rho}
\left(\JJ\times\BB - \nab p+\nab\cdot 2\nu\rho\SSSS\right)\\
\label{mom}
T{\DD s\over\DD t}&=&{1\over\rho}\nab\cdot \left(K\nab T +\chit \rho
  T\nab s\right) + 2\nu\SSSS^2+{\mu_0\eta\over\rho}\JJ^2 - \Gamma_{\rm cool}(r),
\label{entr}
\end{eqnarray}
where magnetic field is expressed by $\BB = \nab \times \AAA$, where
$\AAA$ is the vector potential, and $\eta$ is the magnetic
diffusivity.
Furthermore, $\uu$ is the velocity, $\rho$ is the density and
$\gggg=-GM\rr/r^3$ is the gravitational acceleration.
The Coriolis and Lorentz forces are given by $ 2\OO \times \uu$ and
$\JJ\times\BB$, respectively, where $\OO$ is the angular velocity and
$\JJ=\mu_0^{-1}\nab\times\BB$ is the current density.
$\SSSS$ is the traceless rate-of-strain tensor and $\nu$ is the
kinematic
viscosity.
The energy conservation equation is solved in terms of the specific
entropy $s$. We include radiative transport in term of the diffusion
approximation with heat conductivity $K$ and model the unresolved
turbulent transport with $\chit$.
The fluid obeys the ideal gas law, $p=(\gamma-1)\rho e$,
where $\gamma=c_p/c_v=5/3$ is the ratio of specific heats at constant
pressure and constant volume, respectively, and $e=c_v T$ is the internal
energy density, which defines the temperature $T$.
The third and fourth terms on the rhs of \Eq{entr} describe viscous
and Ohmic heating.
The radial cooling function $\Gamma_{\rm
  cool}(r)=\Gamma_0\left(\cs^2-c^2_{s0}(r)/c^2_{s0}(r)\right)$ is defined
such that it cools/heats the corona towards a predefined temperature
profile $c_{s0}(r)$, which is plotted in \Fig{strat}.
As an initial condition we use a stratification profile similar to
that plotted in \Fig{strat} and a weak, random, Gaussian-distributed seed
magnetic field in the convection zone.

We apply periodic boundary conditions in the azimuthal direction.
For the velocity field we use stress-free conditions on all other
boundaries.
For the magnetic field we use a perfect conductor boundary condition at
the $r=0.7\, R$ and both $\theta$ boundaries; at the $r=1.6\, R$ boundary be
apply a vertical-field condition.
The heat flux through the lower ($r=0.7\, R$) boundary is held constant and
the temperature at the top is fixed.

We employ the 
{\sc Pencil Code}\footnote{\texttt{http://pencil-code.googlecode.com}},
which uses sixth-order centered finite differences in space and 
a third-order Runge-Kutta scheme in time;
see \cite{MTBM09} for the extension to spherical coordinates. 
\section{Results}
For this work we consider a single simulation, which is defined by
the following properties:
the Coriolis number $\Co=2\Omega/\urms\kf$ is 14, the Reynolds number
$\Rey=\urms/\nu\kf$ is $20$, and the magnetic Prandtl number
$\Pm=\nu/\eta=1$.
Here $\kf$ is the wavenumber of the energy carrying scales and is defined as
$\kf R=21$, and $\urms=\sqrt{3/2\bra{u_r^2+u_{\theta}^2}}$ is the rms 
velocity at radius $r=0.84\, R$.
\begin{figure}[t!]
\begin{center}
\includegraphics[width=0.49\columnwidth]{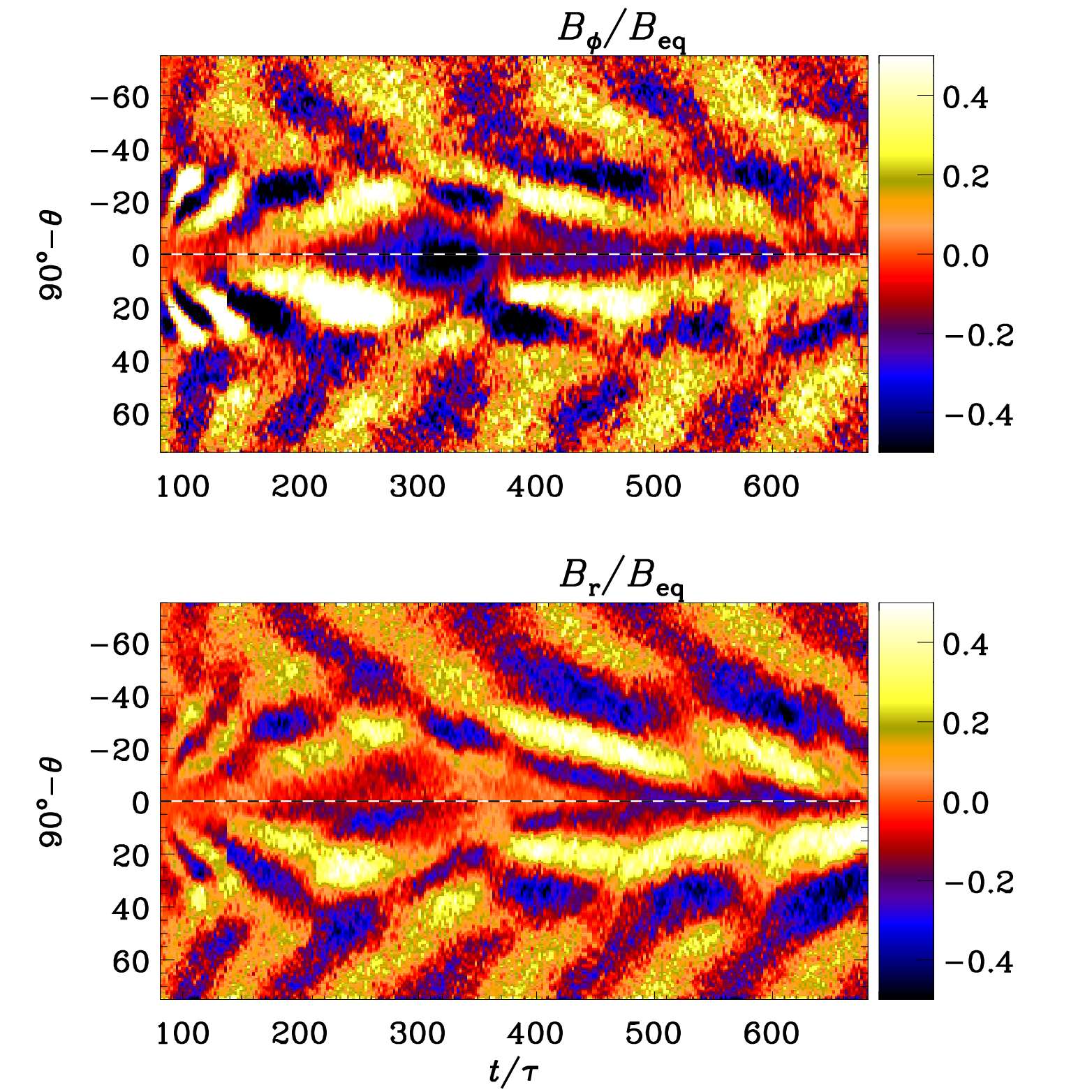}
\includegraphics[width=0.49\columnwidth]{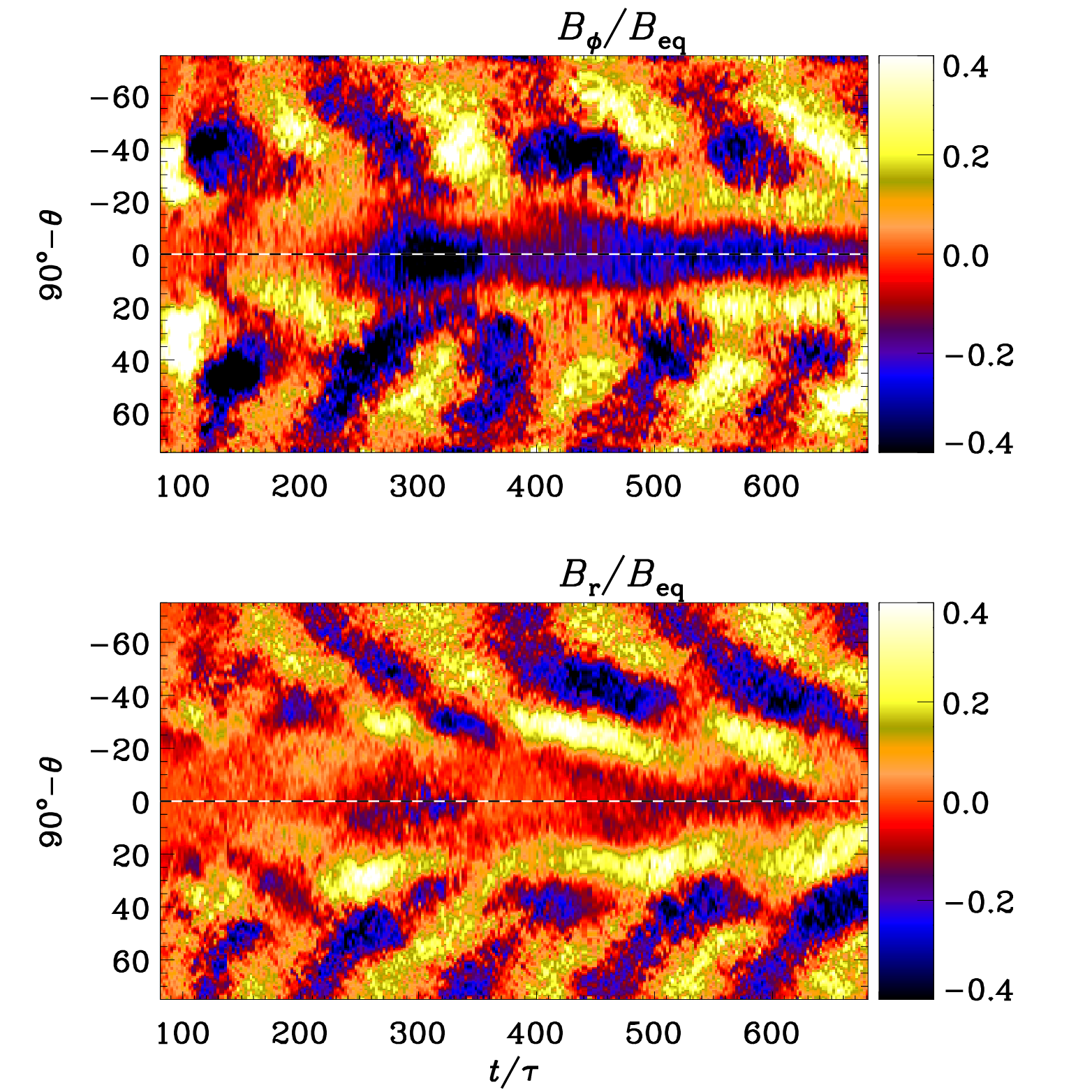}
\end{center}\caption[]{
Variation of the magnetic field in the convection zone.
The azimuthal mean magnetic field $\mean{B_{\phi}}$ and the
radial mean magnetic field $\mean{B_{r}}$ is shown at radius of
$r=0.97\, R$ (left panel)  and 
$r=0.84\, R$ (right).
Dark blue shades represent negative and light yellow positive values.
The dashed horizontal lines show the location of the equator at
$\theta=\pi/2$.
The magnetic field is normalized by the equipartition value $\Beq$.
}
\label{butterfly}
\end{figure}

Below the surface ($r<R$) the domain is convectively unstable.
The resulting convective motions drive a turbulent dynamo and generate a
large--scale magnetic field.
The rms magnetic field $\Brms$ in the convection zone
reaches 0.38 times the equipartition field
$\Beq=\sqrt{\mu_0\bra{\meanrho\urms^2)}_{r=0.84\, R}}$.
In \Fig{butterfly} we plot the azimuthally averaged (in the following
mean) radial and azimuthal magnetic
fields as functions of time in terms of the turnover time $\tau
=\urms\kf$.
Effectively $\tau$ is an eddy turnover time, which varies between a
few minutes and more than a month in the solar convection zone,
depending on the depth.
In the kinematic regime, the magnetic field pattern shows a cyclic poleward
migration close to the surface ($r=0.97\, R$) for the azimuthal
$\mean{B_{\phi}}$ and radial fields $\mean{B_{r}}$ at lower latitudes.
The period is short, only a few tens $t/\tau$.
In the non-linear regime, a new pattern is developing:
equatorward migration starting at high latitudes overcomes
the poleward migration at low latitudes.
This equatorward migration shows a longer period of 100-200 $t/\tau$.
Additionally the pattern is not only visible near the surface, as it
penetrates into the convection zone as shown in the left panel
of \Fig{butterfly}.
The pattern does not look as regular as in \cite{KMB12}, which can be
explained by the interaction of the corona and convection zone.

Initially the interior rotation profile is constant in cylinders
with a positive gradient of angular velocity at low latitudes.
In the hydrodynamic state, affected by rotation, the convective
turbulence re-shapes the rotation profile within the convection zone,
while in the convectively stable coronal layer only viscosity is at
play.
After a dynamically important magnetic field has been generated by the
dynamo action, it influences the rotation as well.
In the corona, where the density is much lower than in the convection
zone, the Lorentz force can dominate over the Coriolis force, see
\Eq{mom}, and have a greater influence on the angular velocity.
In our simulation the velocities in the corona saturate
not before $t/\tau=660$.

\begin{figure}[t!]
\begin{center}
\includegraphics[width=0.40\columnwidth]{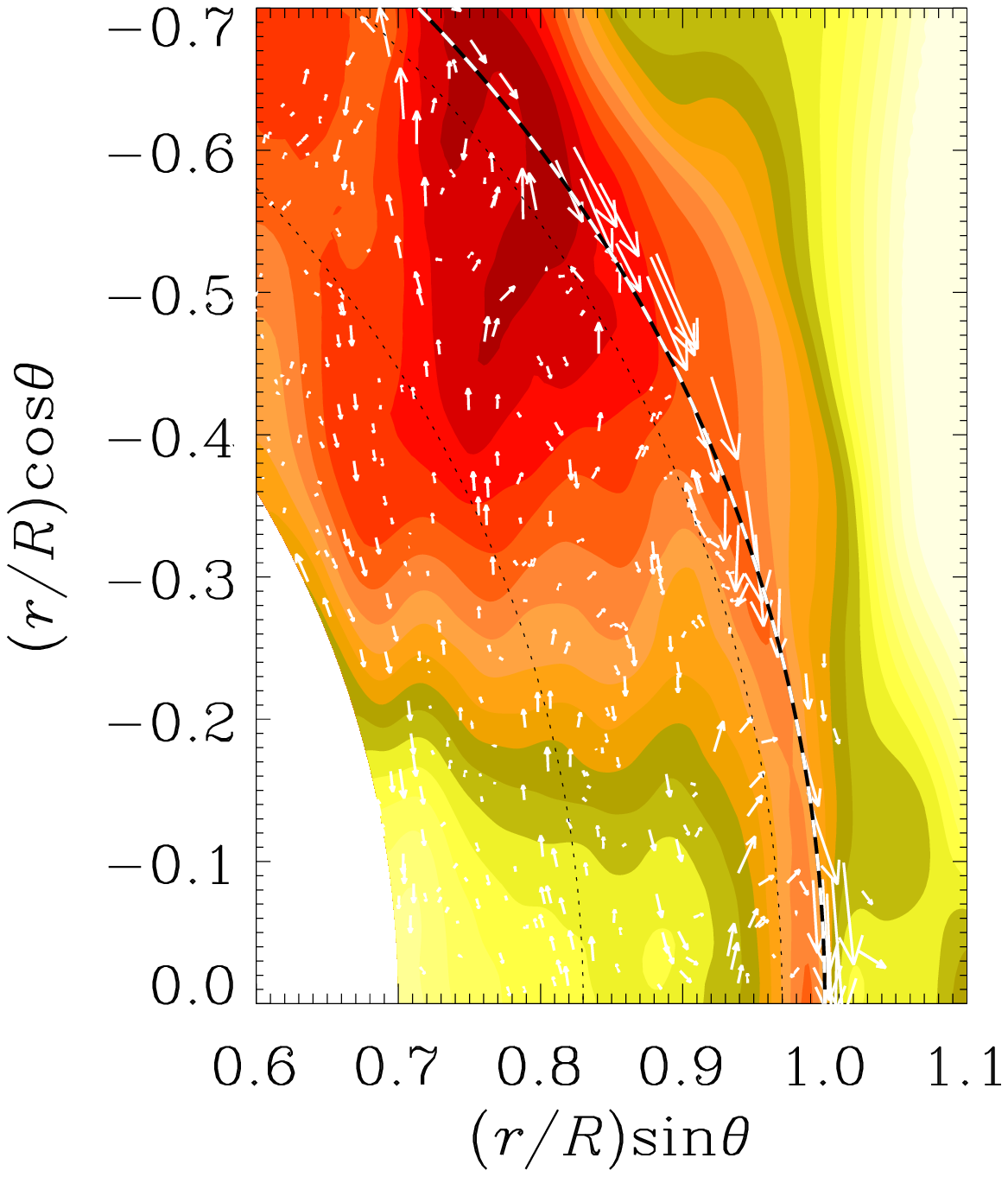}
\includegraphics[width=0.40\columnwidth]{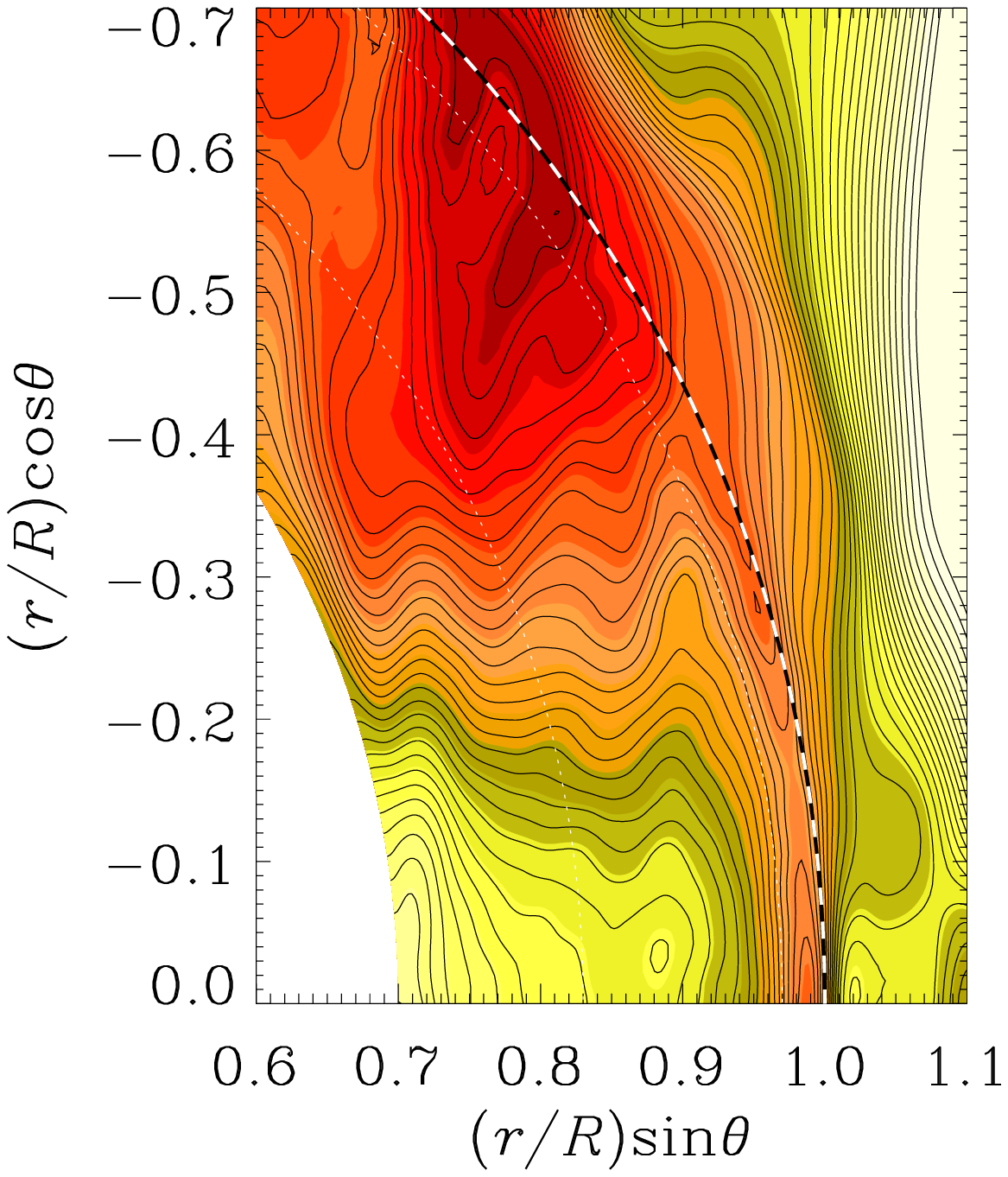}
\end{center}\caption[]{Meridional circulation and differential
  rotation zoomed in the northern hemisphere in the convection
  zone. {\it Left panel}: differential mean rotation profiles
  $\mean{\Omega}(r,\theta)/\Omega_0=\mean{u}_{\phi}/(r\sin{\theta})+1$
  overplotted with the poloidal velocity $\mean{u}_p=\mean{u}_r+\mean{u}_\theta$ in
  white arrows. {\it Right panel}: the same as left, but without the
  poloidal velocity and emphasizing the mean differential rotation
  contours.
The white-black dashed line indicates the surface ($r=R$).
}
\label{meri}
\end{figure}

The differential rotation profile is solar-like, see \Fig{meri}.
The equator rotates faster than the poles and the contours of constant
rotation show a {\it spoke-like} pattern--especially at low latitudes.
This is the first time a spoke-like profile has been found in DNS in 
combination with equatorward migration.
\cite{KMB12}, whose setup is similar to ours in the convection zone,
found that the contours of constant rotation are significantly more
cylindrical at similarly high Coriolis numbers.
This could be due to the interaction between the coronal layer and the
convection zone.
There might be even an indication of a surface shear layer at
latitudes $\theta\leq \pi/3$.
The meridional circulation shows a clockwise pattern--in contrast to
the Sun.
Instead of a poleward flow near the surface, as in the Sun, the flow is
equatorward.
This might be one possible explanation for the equatorward migration
of the magnetic field.
The field would then be dragged along with the fluid motion to the
equator.
But the magnetic field shows an equatorward migration throughout the
convection zone, and the anti-solar meridional circulation seems to be
shallow with a return flow in the middle of the
convection zone.
In our simulation the meridional circulation seems not to have a
connection with the equatorward migration and the generation of the
field.

\begin{figure}[t!]
\begin{center}
\includegraphics[width=0.4\columnwidth]{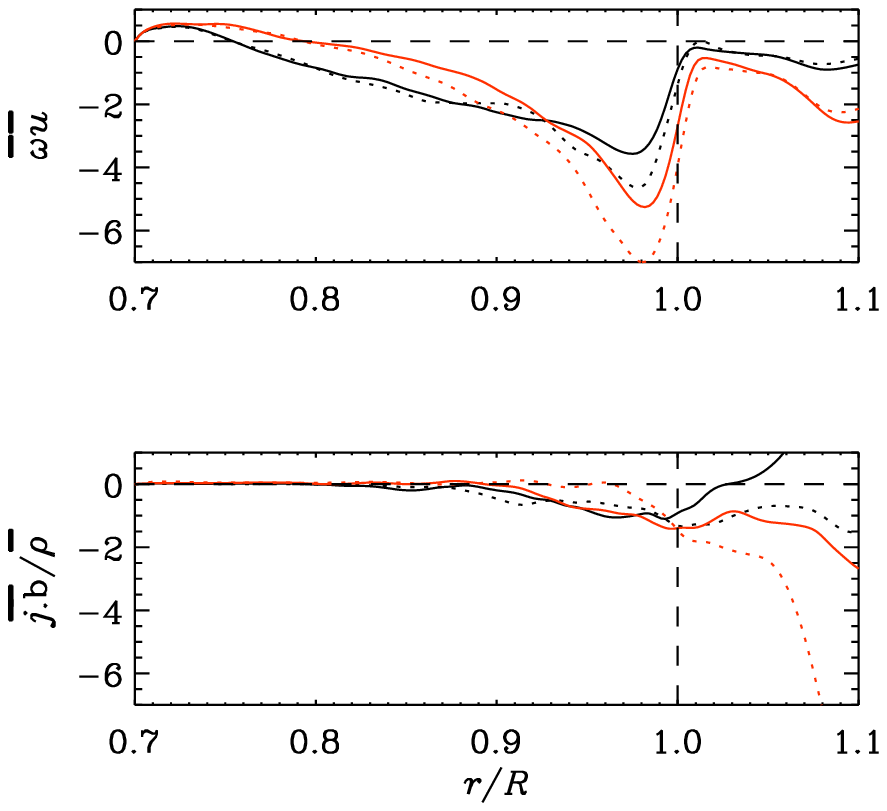}
\includegraphics[width=0.4\columnwidth]{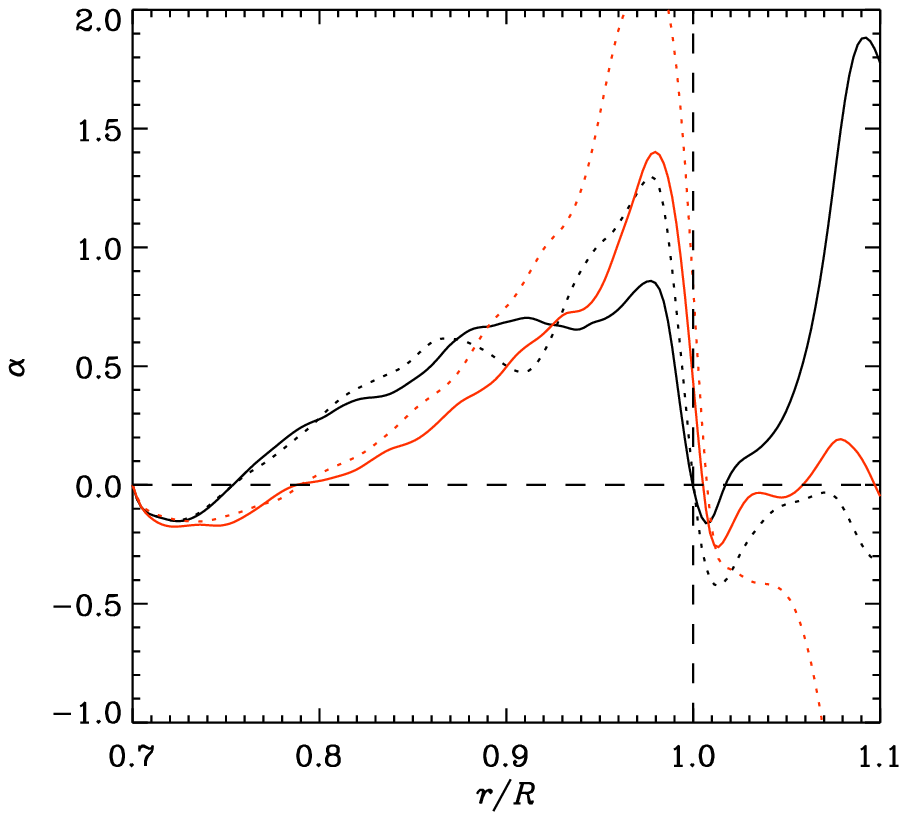}
\end{center}\caption[]{
{\it Left panel}: 
The two contributions to $\alpha$ plotted as functions of the radius $r$. 
Upper left panel: kinetic helicity density $\overline{\oo\cdot\uu}$
for two latitudes: $\theta=70^\circ$ (black line) and $\theta=50^\circ$ (red
line). Lower left panel: current helicity density 
$\overline{\jj\cdot\bb}/\meanrho$
at $\theta=70^\circ$ (black line) and $\theta=50^\circ$ (red line).
The values are normalized by $\urms^2 k_{\rm f}$.
{\it Right panel}: $\alpha=-\onethird\tau_{\rm c}\overline{\oo\cdot\uu} +
\onethird\tau_{\rm c}\overline{\jj\cdot\bb}/\meanrho$
at $\theta=70^\circ$ (black line) and $\theta=50^\circ$ (red line).
The dotted lines indicate the same as above, except in the southern
hemisphere where it has the opposite sign.
$\alpha$ are normalized by $\urms$.
The dashed lines represent the surface ($r=R$) and the zero line.
}
\label{palpha}
\end{figure}

Another possible explanation of the equatorward migration can be the
profile of the effective $\alpha$ which likely generates the mean magnetic
field $\meanBB$.
$\alpha$ has two contributions
\begin{equation}
\alpha=\alpha_{\rm k}+\alpha_{\rm m}=-\onethird\tau_{\rm c}\overline{\oo\cdot\uu}+\onethird\tau_{\rm c}\overline{\jj\cdot\bb}/\meanrho.
\end{equation}
proportional to the kinetic helicity density $\overline{\oo\cdot\uu}$
and the magnetic helicity density $\overline{\jj\cdot\bb}/\meanrho$,
respectively, and where $\tau_{\rm c}$ is a correlation time of the 
turbulence and can be expressed by the eddy turnover time $\tau$.
In \Fig{palpha}, we show the two contribution as functions of radius
$r$ for two latitudes.
In the convection zone $\alpha$ is dominated by the kinetic
contribution and is influenced by $\alpha_{\rm m}$ only close
to the surface.
One reason behind the equatorward migration could be the different values
of $\alpha$ at different latitudes.
We observe a tendency for the $\alpha$ effect to have larger values at
higher latitudes.
The latitudinal profile of $\alpha$ could cause the mean field to 
migrate towards
the equator.
This has to be verified by comparison with simulations which show a
poleward migration.

\section{Conclusions}
In summary, we use a two-layer model that combines
self-consistent convective dynamo action with a coronal layer,
resulting in equatorward migration of the mean magnetic field.
A solar-like rotation profile with spoke-like contours
at low latitudes is generated. This is different from recent results
of \cite{KMB12} with a similar model, but without corona.
In forthcoming work, we will study the mean magnetic field structure
and differential rotation for different rotation rates and investigate 
the cause of
equatorward migration in more detail.


\begin{thebibliography}{
}
\bibitem[Brown et al.(2010)]{BBBMT10}
Brown, B. P., Browning, M. K., Brun, A. S., Miesch, M. S., \& 
Toomre, J.\yapj{2010}{711}{424} 

\bibitem[Brown et al.(2011)]{BMBBT11}
Brown, B. P., Miesch, M. S., Browning, M. K., Brun, A. S. \& 
Toomre, J.\yapj{2011}{731}{69} 

\bibitem[Brun et al.(2004)]{BMT04}
Brun, A. S., Miesch, M. S., \& Toomre, J.\yapj{2004}{614}{1073} 

\bibitem[Dikpati \& Charbonneau(1999)]{DC99}
Dikpati, M., \& Charbonneau, P.\yapj{1999}{518}{508}

\bibitem[Ghizaru et al.(2010)]{GCS10}
Ghizaru, M., Charbonneau, P., \& Smolarkiewicz, P. K. 
\yapjl{2010}{715}{L133} 

\bibitem[Gilman(1983)]{G83}
Gilman, P. A.\yapjs{1983}{53}{243} 

\bibitem[K\"apyl\"a et al.(2006)]{KKT06}
K\"apyl\"a, P. J., Korpi, M. J. \& Tuominen, I. 2006, {AN,} {327}, 884 

\bibitem[K\"apyl\"a et al.(2010)]{KKBMT10}
K\"apyl\"a, P. J., Korpi, M. J., Brandenburg, A., Mitra, D. \& Tavakol, R. 2010, {AN,} {331}, 73 

\bibitem[K\"apyl\"a et al.(2011)]{KKGBC11}
K\"apyl\"a, P. J., Korpi, M. J., Guerrero, G., Brandenburg, A. \&
Chatterjee, P. 2011, {A\&A,} {531}, A162

\bibitem[K\"apyl\"a et al.(2011)]{KMB11}
K\"apyl\"a, P. J., Mantere, M. J., \& Brandenburg, A. \yan{2011}{332}{883} 

\bibitem[K\"apyl\"a et al.(2012)]{KMB12}
K\"apyl\"a, P. J., Mantere, M. J., \& Brandenburg, A. \yapjl{2012}{755}{L22} 

\bibitem[Kitchatinov \& R\"udiger(1999)]{KR99}
Kitchatinov, L. L. \& R\"udiger, G. \yana {1999}{344}{911}

\bibitem[Kitchatinov \& Olemskoy(2012)]{KO12}
Kitchatinov, L. L. \& Olemskoy, S. V.\ysph{2012}{276}{3}

\bibitem[Miesch et al.(2006)]{MBT06}
Miesch, M. S., Brun, A. S. \& Toomre, J.\yapj{2006}{641}{618} 

\bibitem[Mitra et al.(2009)]{MTBM09}
Mitra, D., Tavakol, R., Brandenburg, A., \& Moss, D.\yapj{2009}{697}{923}

\bibitem[Warnecke \& Brandenburg(2010)]{WB10}
Warnecke, J., \& Brandenburg, A.\yana{2010}{523}{A19}

\bibitem[Warnecke et al.(2011)]{WBM11}
Warnecke, J., Brandenburg, A., \& Mitra, D.\yana{2011}{534}{A11}

\bibitem[Warnecke et al.(2012a)]{WBM12}
Warnecke, J., Brandenburg, A., \& Mitra, D.\yswsc{2012a}{2}{A11}

\bibitem[Warnecke et al.(2012b)]{WKMB12}
Warnecke, J., K\"apyl\"a, P. J., Mantere, M. J., \& Brandenburg, A.\ysph{2012b}{280}{299}

\end{thebibliography}
\end{document}